\documentclass
[%
reprint,
superscriptaddress,
showpacs,
amsmath,amssymb,
aps,
pre,
floatfix,
]
{revtex4-2}

\usepackage{textcomp}

\usepackage{xcolor}

\usepackage[hidelinks]{hyperref}

\usepackage{epstopdf}
\usepackage[T1]{fontenc}
\usepackage{graphicx}
\usepackage{comment}
\usepackage{siunitx}
\usepackage{multirow}
\usepackage{booktabs}
\usepackage{algorithm}
\usepackage{algorithmic}

\newcommand\citeay[1]{%
	\citet{#1}}

\newcommand{\revision}[1]{\textcolor{black}{#1}}

\usepackage{xcolor}
\usepackage[utf8]{inputenc}
\usepackage[english]{babel}
\usepackage{graphicx}
\usepackage{comment}
\usepackage{siunitx}
\usepackage{multirow}
\usepackage{booktabs}

\usepackage{lipsum}

\begin{document}

\title{Approach to predicting extreme events in time series of chaotic dynamical systems using machine learning techniques}

\author{Alexandre C. Andreani}
\email{alexandre.c.andreani@gmail.com}
\affiliation{Institute of Science and Technology, Universidade Federal de S\~ao Paulo, 12247-014, S\~ao Jos\'e dos Campos, S\~ao Paulo, Brazil}
\author{Bruno R. R. Boaretto}
\email{bruno.boaretto@unifesp.br}
\affiliation{Institute of Science and Technology, Universidade Federal de S\~ao Paulo, 12247-014, S\~ao Jos\'e dos Campos, S\~ao Paulo, Brazil}
\affiliation{Department of Physics, Universitat Politecnica de Catalunya, 08222, Terrassa, Barcelona, Spain}

\author{Elbert E. N. Macau} 
\email{elbert.macau@unifesp.br}
\affiliation{Institute of Science and Technology, Universidade Federal de S\~ao Paulo, 12247-014, S\~ao Jos\'e dos Campos, S\~ao Paulo, Brazil}


\begin{abstract}

This work proposes an innovative approach using machine learning to predict extreme events in time series of chaotic dynamical systems. The research focuses on the time series of the Hénon map, a two-dimensional model known for its chaotic behavior. The method consists of identifying time windows that anticipate extreme events, using convolutional neural networks to classify the system states. By reconstructing attractors and classifying (normal and transitional) regimes, the model shows high accuracy in predicting normal regimes, although forecasting transitional regimes remains challenging, particularly for longer intervals and rarer events. The method presents a result above $80\%$ of success for predicting the transition regime up to 3 steps before the occurrence of the extreme event. Despite limitations posed by the chaotic nature of the system, the approach opens avenues for further exploration of alternative neural network architectures and broader datasets to enhance forecasting capabilities.
\end{abstract}


\maketitle

\section{Introduction}

Extreme events are rare and significant deviations from average behavior or expected patterns in data, often associated with natural disasters \cite{nearing2024}, market crashes \cite{chang2024}, and critical system failures \cite{tiedmann2024}, as extensively discussed in \citeay{ghil2011}. Understanding and predicting these events is vital for developing mitigation strategies and enhancing system resilience across various domains, including climate science, finance, and engineering \cite{pavithran2021}.

The study of extreme events in dynamical systems has gained considerable attention, with researchers exploring their emergence, predictability, and statistical properties. For instance, \citeay{durairaj2023} investigates the emergence of extreme events in quasi-periodic oscillators, while \citeay{altmann2006} examines the role of threshold positioning in defining extreme events. \citeay{yuan2024} provides insights into the fundamental limits of predicting these phenomena, and \citeay{schweigler2011} analyzes the clustering and recurrence of extreme events. These works collectively underscore the importance of understanding extreme events within the framework of dynamical systems.

In this study, we propose a novel method for predicting extreme events in time series of chaotic systems. Our approach involves training a convolutional neural network (CNN) to classify segments of time series preceding extreme events (transition regimes) from those associated with normal behavior (normal regimes). CNNs are particularly well-suited for this task due to their ability to extract implicit features from time series and capture hidden signatures of transitions to extreme events \cite{Mehrabbeik2025}. For instance, \citeay{wang2024a} used CNN to identify extreme events in southern China based on large-scale atmospheric circulation patterns, the CNN was able to correctly identify about 96\% of the offered extreme events.

Machine learning has demonstrated exceptional promise in forecasting nonlinear dynamical systems, including those exhibiting extreme events. For example, \cite{sun2023} utilizes LSTMs to analyze the five-degree-of-freedom Duffing oscillator system, while \cite{guo2020} employs deep learning to predict the quasi-cyclical climate phenomenon El Niño. Additionally, \cite{teng2019} combines CNNs, LSTMs, and deep neural networks (DNNs) to forecast the behavior of the two-dimensional damped harmonic oscillator. These advancements highlight the growing synergy between machine learning techniques and the study of dynamical systems.

To test our method, we analyze the Hénon map, a two-dimensional discrete-time dynamical system introduced by Michel Hénon in 1976 as a simplified model of the Lorenz system’s Poincaré section \cite{henon1976}. The Hénon map exhibits chaotic behavior for specific parameter values, which can be considered as a reference for studying extreme events in dynamical systems \cite{lellep2020}. Although the extreme events in the Hénon map do not pertain to natural phenomena, they are analogous to those in other dynamical systems, characterized by rare, significant deviations from standard behavior driven by sensitivity to initial conditions \cite{mishra2020}. These extreme events can manifest as abrupt transitions between attractor regions or as large deviations in system variables.

The study of extreme events in dynamical systems has broad relevance, from predicting natural disasters such as earthquakes, tsunamis, and hurricanes \cite{tinti2021, setyonegoro2024, lopez-reyes2023}, to understanding financial crises and sudden market crashes \cite{strogatz2015}. Insights from such analyses contribute to designing more robust systems capable of withstanding disruptions and failures \cite{kaveh2020}.

CNNs have been successfully applied to time series analysis across various fields. For example, \cite{abumohsen2024} combined CNNs with LSTMs and Random Forest for solar power forecasting, while \cite{salim2024} used CNN-LSTM models with image representations to predict gold prices. Similarly, \cite{lu2024} leveraged CNNs and LSTMs to predict chaotic time series. These studies show the versatility and effectiveness of CNNs in capturing complex temporal patterns, making them a very well craft tool for identifying precursors to extreme events.

The structure of the article is as follows: Section \ref{sec:methods} details the methodology, including the identification of extreme events in the Hénon map, the dataset generation process, and the machine learning approach employed. Section \ref{sec:results} presents the results, and Section \ref{sec:conc} concludes the study with a discussion of the findings and potential future directions.

\section{Methods}\label{sec:methods}

\subsection{Hénon map and routes to extreme events} \label{sec:henon}

\begin{figure*}[htb!]
	\centering
	\includegraphics[width=1.92\columnwidth]{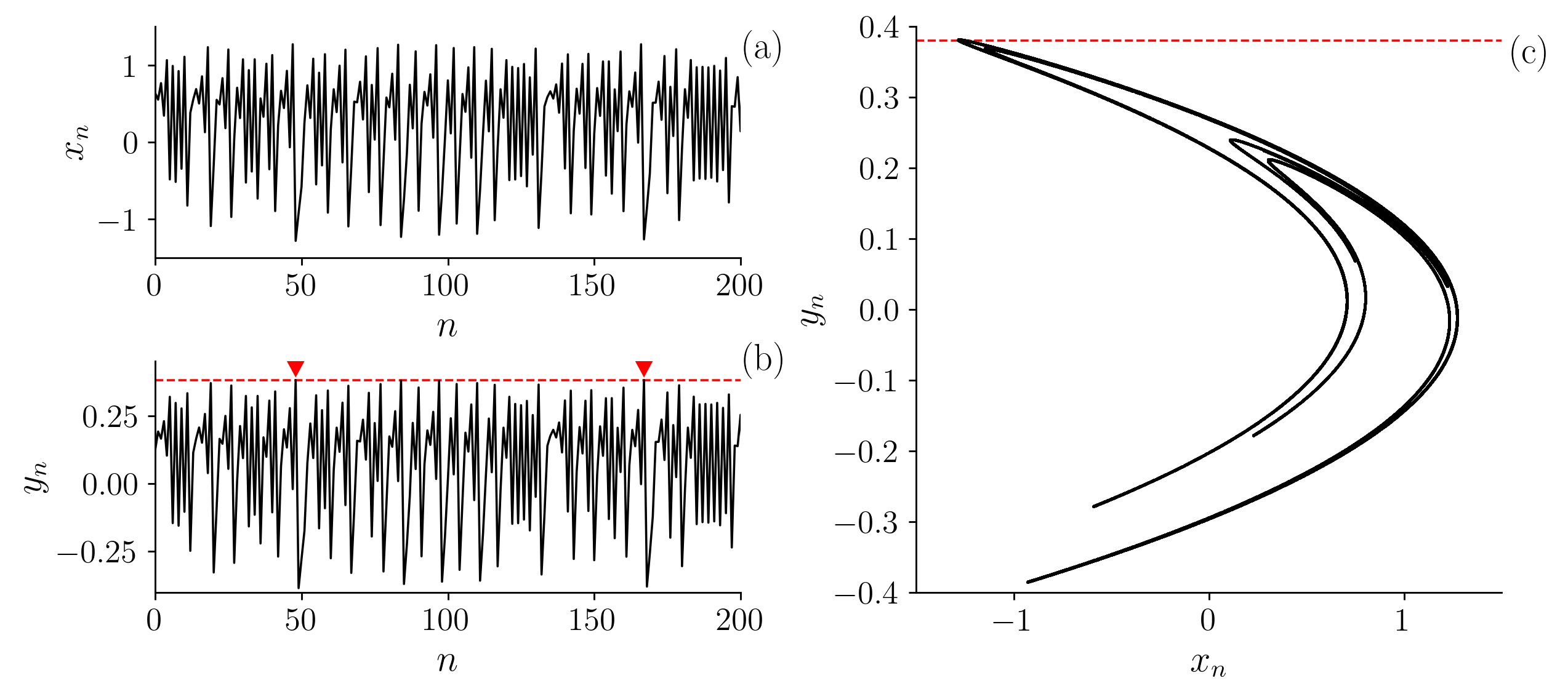}
	\caption{\label{fig:henonseries}Time series of the Hénon map for parameters $a = 1.4$ and $b = 0.3$. Panel (a) shows the evolution of $x_n$, and panel (b) presents the corresponding $y_n$. Panel (c) illustrates the Hénon attractor in the $x$-$y$ phase space. The dashed line in panels (b) and (c), $y^* = 0.38$, represents the threshold used to identify extreme events while the two red triangles in (b) correspond to two extreme-event occurrences. 
}
\end{figure*}

The Hénon map is defined by the following equations \cite{alligood1996}:
\begin{eqnarray}
    	x_{n+1} & = & 1-ax_n^2+y_n, \\
		y_{n+1} & = & bx_n,
\end{eqnarray}
where $x_n$ and $y_n$ are the coordinates of the point in the plane at time $n$, and $a$ and $b$ are parameters that control the behavior of the system. The variable $a$ is critical in determining the amount of nonlinearity and the complexity of chaotic behavior, while $b$ controls the contraction in the $y$ direction. Figure \ref{fig:henonseries} displays the time evolution of $x_n$ (a) and $y_n$ (b) for $a = 1.4$ and $b = 0.3$, which are classical values often used to demonstrate chaotic behavior and are kept fixed throughout this study. Panel (c) shows the chaotic attractor of the Hénon map, where the points in the $x$-$y$ phase space form a characteristic structure with dense regions and a complex, fractal-like pattern, illustrating the underlying chaotic dynamics of the system. 

\begin{figure}[htb!]
    \centering
    \includegraphics[width=0.95\linewidth]{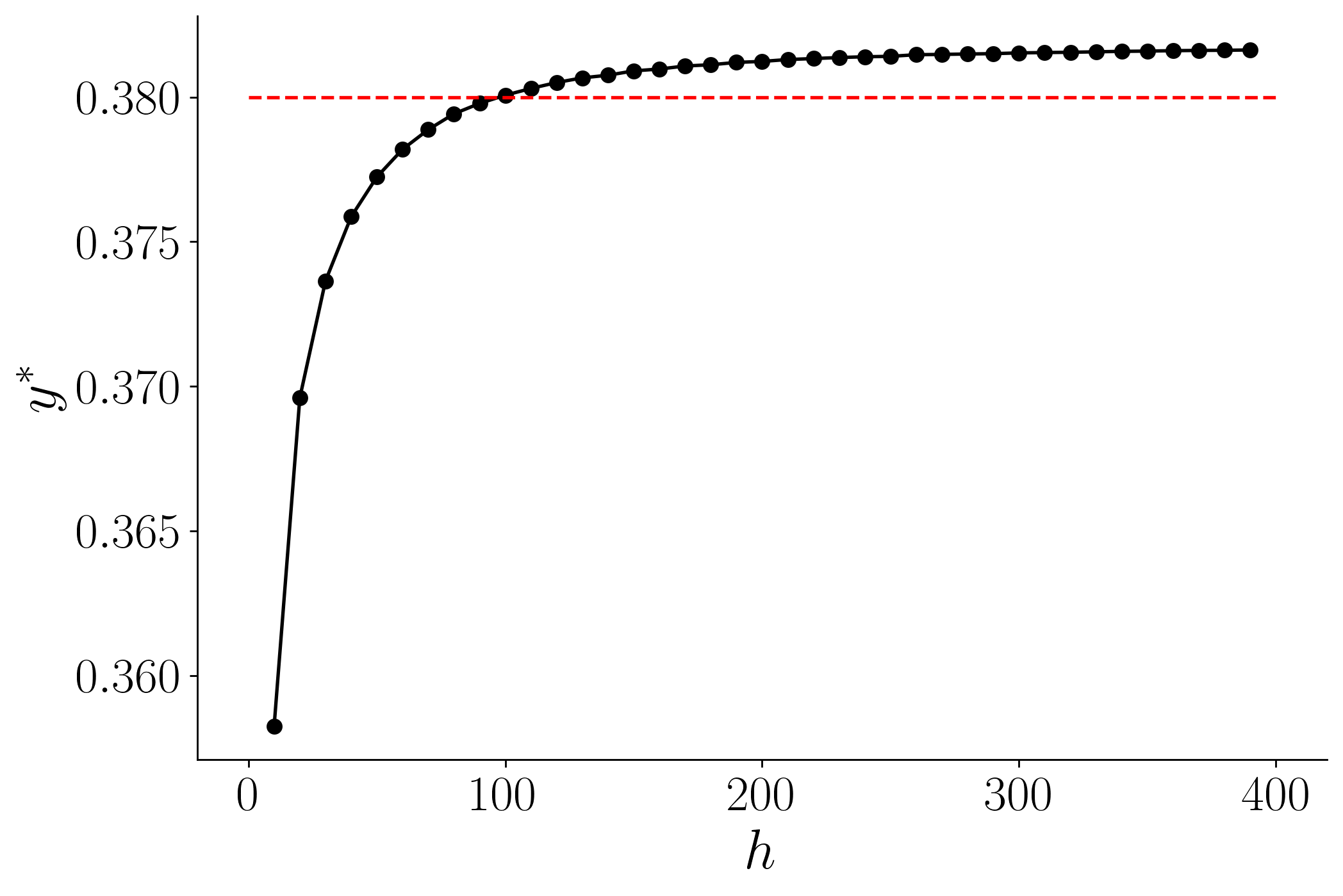}
    \caption{Threshold \(y^*\) as a function of the segment length \(h\) for a fixed value of \(c=8\). The threshold \(y^*\) is determined as \(y^* = \mu + c\sigma\), where \(\mu\) is the mean value of the maxima points over all segments, and \(\sigma\) is the corresponding standard deviation. As the segment length \(h\) increases, the probability of access greater maximum increases resulting in a higher threshold \(y^*\). The red dashed line represents the established threshold at $y^* = 0.38$.}
    \label{fig:limiar}
\end{figure}

From a general perspective, the definition of extreme events lacks a strict consensus, largely due to their diverse interpretations in the literature. A more detailed discussion on it can be found in \citeay{Broska2020}. As outlined in \citeay{mishra2020}, events occurring in less than $1\%$ of the recorded cases are often classified as extreme \cite{Joseph2024}. Following the methodologies proposed by \citeay{mishra2020} and \citeay{ray2019}, a fixed threshold for identifying extreme events is determined as follows: 

First, we analyze a long time series of $y_n$, with a fixed length $L=400\,000$, divided into $m$ non-overlapping segments of equal length $h$. For each segment, we extract the maximum point, called $y_{\mathrm{max},j}$ for $j=1,\,\cdots,\,m$. With this information, we evaluate the mean value over maxima points $\mu = \langle y_{\mathrm{max},j}\rangle $, and its standard deviation $\sigma(y_{\mathrm{max},j})$. The threshold $y^*$ is then defined, by studying $y^* = \mu + c\sigma$ where $c$ is a tunable parameter that determines the rarity of the extreme events. We set $c = 8$, aligning with methodologies in the literature that typically consider values between $4$ and $8$, ensuring robust identification of significant extreme events while minimizing false positives. Figure \ref{fig:limiar} illustrates the value of threshold $y^*$ as a function of the length of the segments $h$ for a fixed value of $c=8$. We observe that as the length of the segments $h$ increases, the probability of access greater maximum increases resulting in a higher threshold \(y^*\).
%

Using this method, we establish the threshold at $y^* = 0.38$ (red dashed line), corresponding to events occurring approximately once every 100 points. In Fig.~\ref{fig:henonseries}, this threshold is illustrated as a red dashed line in panels (b) and (c), with two occurrences highlighted in panel (b) as red triangles above the line.

\begin{figure}[htb!]
    \centering
    \includegraphics[width=.95\linewidth]{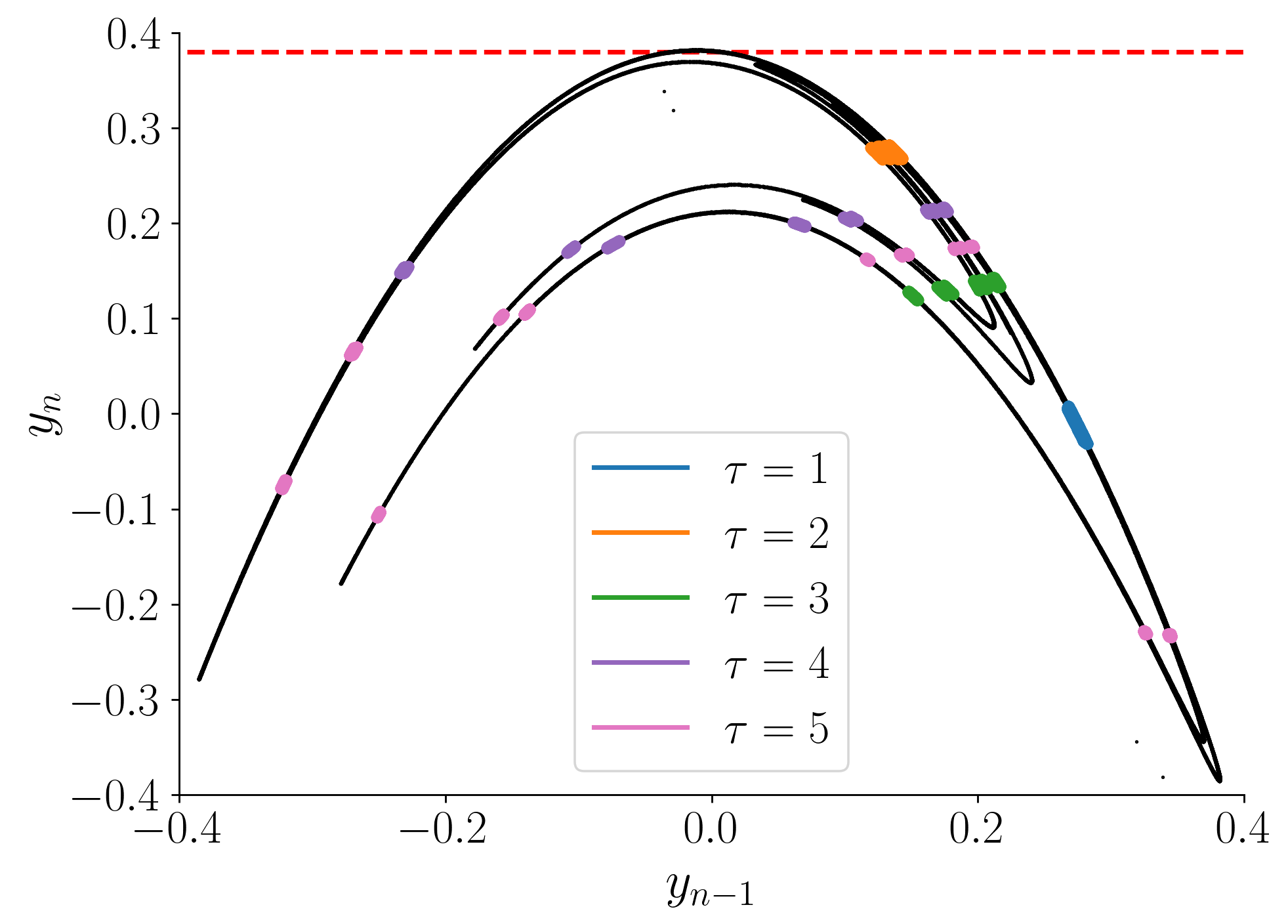}
    \caption{Reconstruction of the attractor for the $y_n$ variable of the Hénon map as a function of $y_{n-1}$. The color code presented over the projection of the attractor represents the points that antecede an extreme event with an interval equal to $\tau$. The dashed line in $y^* = 0.38$, represents the threshold used to identify extreme events.}
    \label{fig:reco}
\end{figure}

Although the Hénon map is a two-dimensional system composed of two-time series ($x_n$ and $y_n$), real-world systems typically provide access to only a single observable. To address this limitation and enhance the complexity of our analysis, we focus exclusively on the time series of the variable $y_n$. To maximize the information extracted from this single time series, we reconstruct the attractor using a time-delay embedding \cite{kantz2003nonlinear}. The reconstruction of the attractor is presented in Fig.~\ref{fig:reco}, where $y_n$ is plotted as a function of $y_{n-1}$.

This reconstruction shows that a single time series can provide significant insights into the system's dynamics, though the choice of embedding dimension and delay depends on the specific system under study \cite{kantz2003nonlinear}. By selecting a fixed threshold of $y^* = 0.38$, as discussed in the previous section and shown as a red dashed line in Figs.~\ref{fig:henonseries}(c) and \ref{fig:reco}, we identify points in the reconstructed attractor that precede an extreme event by $\tau$ steps. These points, highlighted through the color code in Fig.~\ref{fig:reco}, illustrate the complexity of anticipating extreme events. For increasing values of $\tau$, the points become more dispersed, forming distinct clusters in the attractor's phase space. For instance, at $\tau = 1$, a single cluster is evident, whereas at $\tau = 4$, at least six distinct clusters emerge. This growing dispersion reflects the increasing complexity of the system’s dynamics, emphasizing the challenges associated with predicting extreme events as $\tau$ grows.

\subsection{Dataset generation}\label{sec:dataset}

The first step of our method is to generate the dataset to iterate a long time series of the Hénon map. For this, we analyze the time series of $400\,000$ iterations, considering the parameters $a = 1.4$ and $b = 0.3$, and initial conditions $x = 0.1$, $y = 0.3$. To avoid transient effects we discard the first initial $500$ iterations. We have to mention that qualitative results are expected for distinct initial conditions. 

After generating the time series, as a second step, inspired by the work of \citeay{lellep2020}, we manually classify non-overlapping windows of length $W$ into two distinct categories: transition segments (TR) and normal segments (N). TR segments are defined as windows that precede an extreme event by a specific delay, $\tau$, while N segments are windows located sufficiently far from any extreme event. This classification becomes straightforward with the use of a well-defined threshold, enabling precise identification of extreme events. Once an extreme event is detected, a TR segment can be extracted as the window immediately preceding it by $\tau$ steps. In contrast, N segments are windows that do not correspond to transitions or contain any extreme event.

\begin{figure*}
    \centering
    \includegraphics[width=\linewidth]{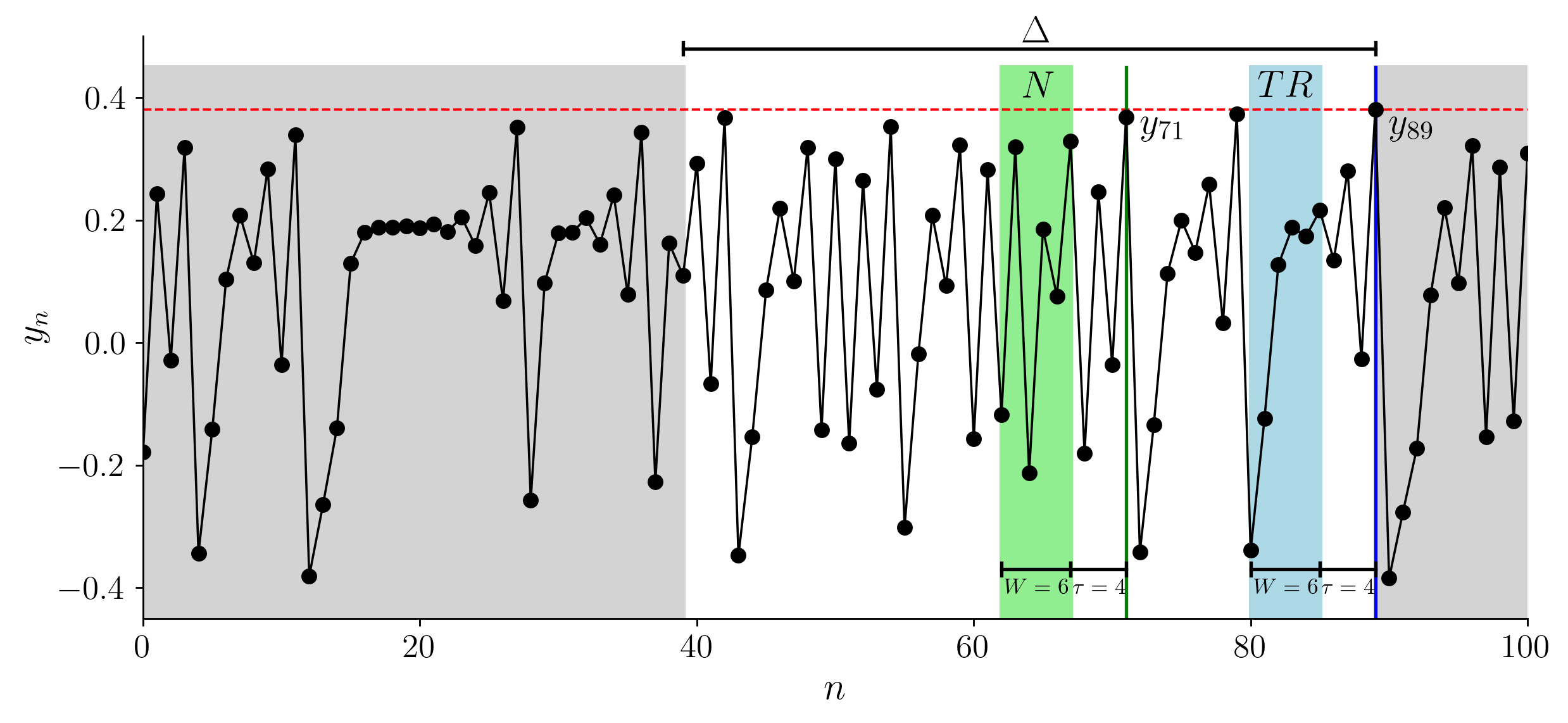}
    \caption{Illustration of the segmentation process for the time series $y_n$, showing the classification of transition (TR) and normal (N) segments. In this example, we use segments of length $W=6$ that precede a specific point with a delay $\tau=4$. An extreme event is detected at $y_{89}$. The segment $\tau$ steps before the extreme event is classified as a transition segment (blue rectangle). Conversely, $y_{71}$, which does not correspond to an extreme event, is used to extract a normal segment (green rectangle). The near-extreme events region defined by the interval $\Delta$, where segments are considered for analysis, is shown in white, while segments outside this region are shaded in gray.}
    \label{fig:segmentation}
\end{figure*}

Figure \ref{fig:segmentation} illustrates the extraction of two windows of the system, using $W=6$ and $\tau=4$ as an example. The time series $y_n$ is shown, where an extreme event is detected at $y_{89}$. A transition segment is extracted $\tau$ steps before the extreme event, highlighted by a blue rectangle. Similarly, a normal segment is identified at $y_{71}$, which does not correspond to an extreme event; the segment $\tau$ steps before is marked by a green rectangle.

In our approach, it is crucial to emphasize that the extreme events themselves are excluded from the analysis; only TR and N segments are considered. Typically, sequences of N segments are interrupted by TR segments immediately preceding extreme events. This segmentation ensures an inherent imbalance in the dataset. To address this challenge, we focus exclusively on segments within a specific proximity to extreme events—a region referred to as the near-extreme events region, defined as a time interval of point called $\Delta$. In the example of Fig. \ref{fig:segmentation}, $\Delta=50$, corresponds to a region highlighted in white, while segments outside this region are shaded in gray rectangles.

\begin{figure}[h!]
\centering
	\includegraphics[width=.95\columnwidth]{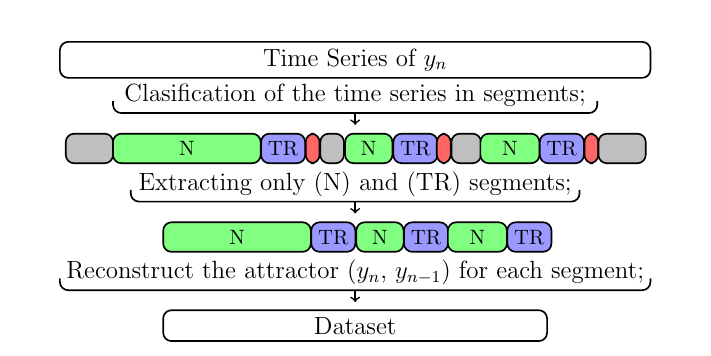}
	\caption{Summary of the method to generate the dataset: after the generation of a time series, the series is classified into windows of interest. The green and blue rectangles represent normal (N) and transition (TR) segments, respectively, which are the states of interest. The red rectangle represents the region of extreme event occurrences, while the gray rectangles correspond to regions of the time series that are far from the next extreme events. Only normal and transition segments are extracted, where each segment is an ordered time series of length $W$. For each segment, we reconstruct the attractor, generating a 2-dimensional array composed of ($y_n$, $y_{n-1}$), which represents a sample in the dataset.}
    \label{fig:regimes}
\end{figure}

Figure \ref{fig:regimes} summarizes the method to generate the dataset: after the generation of a time series, we classify the time series into windows, where we are only interested in the N (normal) and TR (transition) segments. It is important to note that the distinction in the width of the rectangles illustrates the dataset's imbalance, but all segments are composed of a unique, originally ordered time series of length $W$. After extracting only the normal and transition segments, we reconstruct the attractor for each segment. This 2-dimensional array of length $W$ composed of ($y_n$, $y_{n-1}$), represents one sample in the dataset. 

It is also important to note that $\tau$, $W$, and $\Delta$ are free parameters that can be varied during the study. Increasing $\tau$ places TR segments farther from the occurrence of extreme events, potentially allowing a more detailed exploration of pre-event dynamics, but also making the transitions harder to anticipate. Similarly, increasing $W$ reduces the total number of samples available in the dataset, as longer windows require more data to populate each category. For $\Delta$, decreasing its value makes the near-extreme event's region smaller, focusing the analysis closer to extreme events. While this can highlight immediate pre-event behavior, it also limits the temporal range of segments and reduces the ability to observe longer-term dynamics leading to extreme events. These trade-offs must be carefully considered when selecting appropriate values for $\tau$, $W$, and $\Delta$ based on the specific goals of the analysis.

Due to the chaotic nature of the system, extreme events can occur at intervals of varying lengths, including very short ones. In such cases, it becomes impossible to extract N and TR segments for inclusion in the dataset. To address this limitation, we introduce the concept of a Regime minimum size, which defines the minimum allowable interval between two extreme events for analysis. Table~\ref{tab:dataset} presents the details of the dataset using the default parameters of this study: window size $W=50$, delay $\tau=1$, and near-extreme event region $\Delta=300$. The first column specifies the minimum interval size between extreme events. The second column shows the average time between extreme events, which naturally increases as larger minimum intervals are imposed. The third and fourth columns provide the number of N and TR samples, respectively. The fifth column reports the number of TR samples excluded due to the minimum interval constraint. Finally, the last column indicates the total number of samples in the dataset as a function of the specified minimum interval. In this study, we fix the minimum interval size to 50 (highlighted), as the average time between extreme events in the second column is greater than 100 points per event, which aligns with the validity criteria outlined in \citeay{ray2019}.
\begin{table}[ht]
\caption{Dataset characteristics using the default parameters of this study ($W=50$, $\tau=1$, $\Delta=300$). The table specifies the minimum interval size between extreme events, average time between events, counts of N and TR samples, discarded TR samples, and the total dataset size.\label{tab:dataset}}

\setlength{\tabcolsep}{1.8pt} 

    \centering
    \begin{tabular}{c c c c c c}\\[-2pt]
        \hline\\
        \textbf{Regime} & \textbf{Average} & \textbf{N} & \textbf{TR} &  \textbf{Discarded} & \textbf{Total} \\ 
        \textbf{size} & \textbf{size} & \textbf{samples} & \textbf{samples} &  \textbf{samples} & \textbf{samples} \\\\ 
        \hline\\[-2pt]
        10  & 87   & 250,166 & 3,181 & 82   & 253,347 \\[2pt] 
        20  & 97   & 220,010 & 2,788 & 189  & 222,798 \\[2pt]
        \textbf{50}  & \textbf{126}  & \textbf{148,857} & \textbf{1,915} & \textbf{427}  & \textbf{150,772} \\[2pt]
        100 & 172  & 77,525  & 1,031 & 668  & 78,556  \\[2pt] 
        150 & 496  & 38,937  & 545   & 801  & 39,482  \\[2pt] 
        200 & 255  & 17,356  & 305   & 866  & 17,661  \\[2pt] 
        250 & 287  & 6,270   & 156   & 907  & 6,426   \\[1pt] 
        \botrule
    \end{tabular}
\end{table}



\subsection{Machine learning approach} \label{sec:ml}


Originally designed to process images, CNNs have also proven efficient in classifying time series. They take the ability to extract relevant features from grid-organized data, such as images, and apply it to sequential data, such as time series \cite{Sanju2021}. Convolutional layers apply filters (kernels) that slide through time, extracting local features from different time scales. The pooling layer reduces the dimensionality of the data, adding more data and making the model more resistant to small variations in the time series. The classification of the obtained characteristics is done through interconnected layers that classify the time series into several categories \cite{venkatesan2017}. As advantages, CNNs are efficient, as they share weights in their convolutional layers, reducing the number of parameters to be trained. Furthermore, CNNs can recognize objects in different positions in the image and also learn increasingly complex features as information flows through the layers.

\begin{figure}[h!]
\centering
	\includegraphics[width=.95\columnwidth]{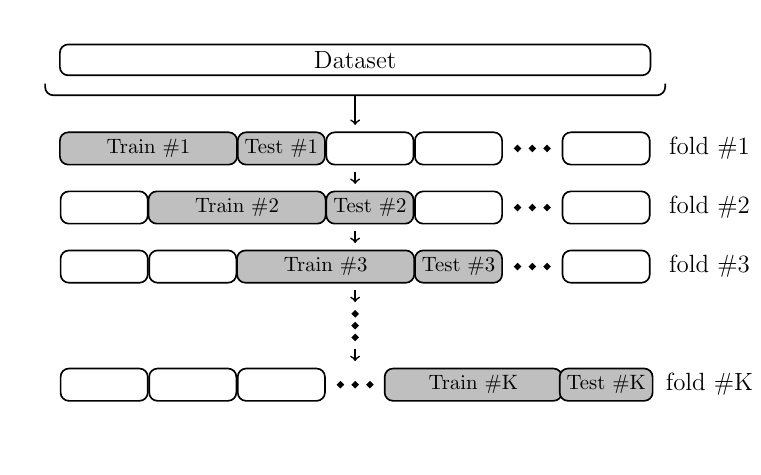}
	\caption{\label{fig:kfold}K-fold-adapted illustration. The test set is next to the training set. For each round of the k-fold, the training and testing set is shifted by the exact size of the test set.}
\end{figure}

After constructing the dataset consisting exclusively of N and TR segments (as described in the previous section), we apply an adapted K-fold cross-validation strategy to partition the data into training and testing sets. This approach divides the dataset into $K$ consecutive folds while preserving the segments' temporal order. At each iteration, one fold is the test set, and the other fold is the training set. Importantly, the test set is always selected as the segment immediately following the training set to maintain temporal consistency and prevent data leakage. After each iteration, the training and test sets are shifted forward by the size of the test set, as illustrated in Fig.~\ref{fig:kfold}. This sequential splitting ensures that the model is evaluated on unseen, temporally ordered segments, making it well-suited for time series data. 


For each fold, the model's performance is quantified by calculating both the accuracy and recall. The accuracy is defined as the ratio of correctly classified segments (TR or N) to the total number of segments in the test set:
\begin{equation}
\text{Accuracy}_k = \frac{\mathrm{TP}_k + \mathrm{TN}_k}{\mathrm{TP}_k + \mathrm{TN}_k + \mathrm{FP}_k + \mathrm{FN}_k}
\end{equation}
where $\mathrm{TP}_k$ is the number of true positives (TR segments correctly classified as TR), $\mathrm{TN}_k$ is the number of true negatives (N segments correctly classified as N), $\mathrm{FP}_k$ is the number of false positives (N segments incorrectly classified as TR), and $\mathrm{FN}_k$ is the number of false negatives (TR segments incorrectly classified as N). 

While accuracy provides an overall measure of performance, it may be misleading in our case due to the dataset's imbalance. Specifically, the dataset contains significantly more N segments than TR segments, meaning that a model biased towards predicting the majority class (N) could achieve a high accuracy without effectively identifying TR segments. To address this limitation, we also compute recall, which is defined as the proportion of actual positive instances the model correctly identified. Mathematically, it is given by:
\begin{equation}
\text{Recall}_k = \frac{\mathrm{TP}_k}{\mathrm{TP}_k + \mathrm{FN}_k}.
\end{equation}
Recall measures the model's ability to capture all relevant instances, making it particularly important in scenarios where missing positive cases is costly, such as in medical diagnoses or fraud detection.
Both the accuracy and recall for each fold are recorded. At the end of the cross-validation process, the overall accuracy and recall of the model are computed as the averages of these $K$ metrics:
\begin{equation}
\text{Average Accuracy} \equiv \alpha = \frac{1}{K} \sum_{k=1}^K \text{Accuracy}_k,
\end{equation}
\begin{equation}
\text{Average Recall} \equiv \beta = \frac{1}{K} \sum_{k=1}^K \text{Recall}_k.
\end{equation}
Combining these metrics provides a more nuanced and reliable evaluation of the model's performance across all folds.

\begin{figure}[h!]
\centering
	\includegraphics[width=.85\columnwidth]{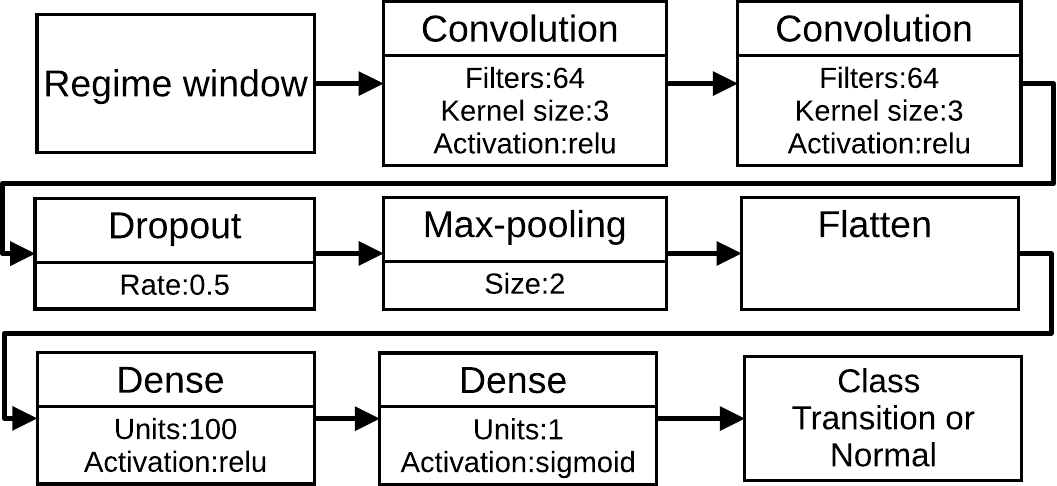}
	\caption{\label{fig:cnn}Configuration of the CNN network, receives as input the attractor reconstruction ($y_n \times y_{n-1}$) of time series of length $W$ for N and TR regimes (See Sec.\ref{sec:dataset}). Two convolution layers, a dropout layer, Max-pooling, Flatten, 2 dense layers, and so on end the exit as a normal regime or transitional regime class }
\end{figure}

The CNN architecture used, illustrated in Fig. \ref{fig:cnn}, presents as input the sequences of states, from the time series, extracted as shown in Fig. \ref{fig:regimes}. The data flow through its different layers, in 8 steps, is as follows:
\begin{algorithm}[H]
\caption{\revision{CNN-Based Classification Procedure}}
\begin{algorithmic}[1]
\STATE \revision{\textbf{Input:} Regime window (a segment of a time series)}
\STATE \revision{\textbf{Step 1:} Apply convolution operations to extract local patterns.}
\STATE \revision{\textbf{Step 2:} Apply dropout to randomly disable neurons during training (regularization).}
\STATE \revision{\textbf{Step 3:} Apply max-pooling with a window size of 2 to reduce spatial dimensionality.}
\STATE \revision{\textbf{Step 4:} Flatten the pooled feature maps into a 1D vector.}
\STATE \revision{\textbf{Step 5:} Pass through a dense layer with 100 neurons using ReLU activation.}
\STATE \revision{\textbf{Step 6:} Pass through a dense layer with 1 neuron using Sigmoid activation (regression output).}
\STATE \revision{\textbf{Output:} Classify the output as ``Transition'' or ``Normal'' based on the predicted value.}
\end{algorithmic}
\end{algorithm}


In summary, this CNN receives segments of the time series, extracts features, reduces the dimensionality and performs a binary classification.

\section{Results} \label{sec:results}

Building on the methods described in Sec.~\ref{sec:methods}, this section presents the outcomes of the proposed approach for classifying segments preceding extreme events in the Hénon map. The analysis considers the Hénon map evolution for $400,000$ steps with parameters $a=1.4$ and $b=0.3$. Extreme events are identified when the $y_n$ variable crosses the threshold $y^*=0.38$. The time series is divided into windows of length $W$, classified as either TR (transition regime) or N (normal), where TR segments anticipate extreme events with a delay $\tau$, and N segments are far from such events. To address the imbalance of the dataset, we focus on samples near extreme events within a region defined by the near-extreme event parameter $\Delta$. Together, $W$, $\tau$, and $\Delta$ constitute the free parameters analyzed in this study. In the following, we evaluate the classification accuracy of the machine learning model using the adapted K-fold cross-validation for $K=5$, varying one parameter at a time while using default values of $W=50$, $\tau=1$, and $\Delta=300$. We call the reader's attention again that all the points of occurrence of extreme events are discarded from this analysis.

\begin{figure*}
    \centering
    \includegraphics[width=0.95\linewidth]{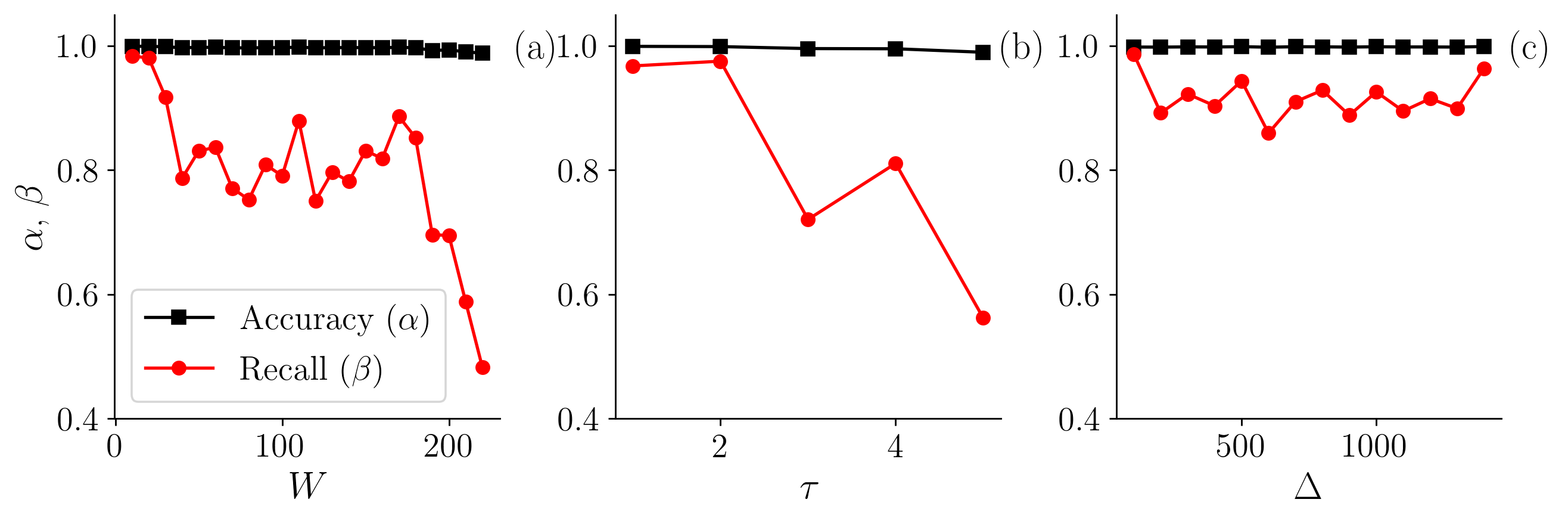}
    \caption{Accuracy ($\alpha$) and recall ($\beta$) as functions of the parameters $W=50$, $\tau=1$, and $\Delta=300$. Each panel presents the variation of $\alpha$ and $\beta$ as a function of one parameter, while the other two are held constant: (a) $W$, (b) $\tau$, and (c) $\Delta$.}
    \label{fig:results_1}
\end{figure*}

The results are presented in Fig.~\ref{fig:results_1}, which shows the accuracy ($\alpha$) and recall ($\beta$) as functions of the parameters $W$, $\tau$, and $\Delta$. Each panel illustrates the effect of varying one parameter while keeping the others fixed. In panel (a) we observe that $\beta$ decreases significantly for $W>180$. Panel (b) shows a decline in $\beta$ as $\tau$ increases, with values dropping below $90\%$ for $\tau > 3$. Panel (c) shows an initial decrease in $\beta$ followed by a plateau, highlighting the sensitivity of detection to variations in $\Delta$. Overall, $\alpha$ remains close to 1 across all parameter ranges, confirming consistent accuracy. The trends observed in $\beta$ provide insights into the parameter dependencies and limitations of the detection framework.

\begin{figure*}
    \centering
    \includegraphics[width=0.95\linewidth]{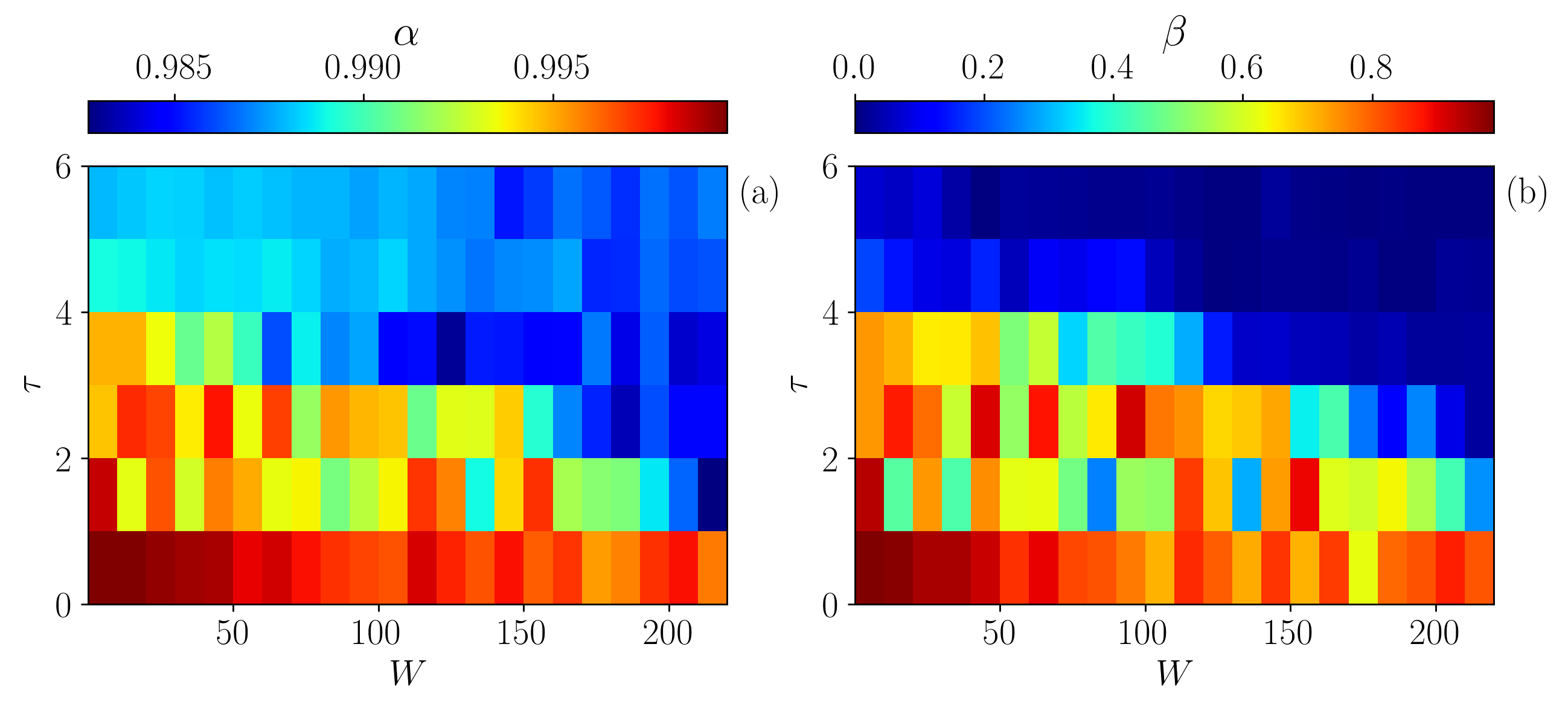}
\caption{Color map of $\alpha$ and $\beta$ in the $W \times \tau$ plane, with values ranging from blue (low) to red (high) for a fixed value of $\Delta = 300$. While both panels provide qualitative insights, they are not to scale.}
    \label{fig:surf}
\end{figure*}

A more general scenario is presented in Figure \ref{fig:surf}, since variations in $\Delta$ have little effect on the results, we fix it at $\Delta = 300$ and focus on analyzing the variations in the $W \times \tau$ plane. Figure \ref{fig:surf} presents a color map ranging from blue (small values of $\alpha$ and $\beta$) to red (high values). We call the attention of the reader that despite qualitative results from both panels, the figures are not in scale. We observe that for $\tau = 1$, both $\alpha$ and $\beta$ remain constant as a function of $W$, showing that the TR (Transition Regime) can be detected just before the transition occurs. For larger $\tau$ values, achieving similar results requires reducing the window size, suggesting that larger windows may blend information from the normal regime. When $\tau > 4$, panel (b) shows extremely low $\beta$ values across the entire range of $W$, indicating that $\tau$ imposes a methodological limitation. 

\section{Conclusion} \label{sec:conc}

    This work has explored the prediction of extreme events in chaotic time series using the Hénon map as a model and convolutional neural networks (CNNs) as a predictive tool. By reconstructing attractors and classifying regimes (normal and transitional), the potential of machine learning in analyzing chaotic systems without full knowledge of the underlying dynamics has been demonstrated. The results have shown high accuracy in classifying transitional regimes, regimes that precede an extreme event occurrence, highlighting the success of the methodology and the possibility of prediction. In addition, the study has shown how window size and prediction interval influence accuracy, emphasizing the trade-offs in sample availability and prediction reliability.

The analysis has revealed that the detection of the TR (Transition Regime) is feasible just before the transition, with results remaining consistent under certain conditions where the accuracy approaches to $100\%$ and the recall above $80\%$. However, larger observation windows have obscured critical information by merging it with the normal regime. Additionally, the methodology has exhibited limitations for increasing parameter values, where meaningful detection has become challenging. This suggests that our methodology has captured the information of the transition not too far and a few steps before the extreme event occurs.


\revision{This study contributes to the ongoing effort of applying machine learning to detect transitions and predict extreme events in chaotic systems. While a fixed threshold was used here to identify segments of the time-series that anticipate extreme events, this approach may be limited in systems with nonstationary behavior or time-varying parameters. Importantly, adaptive thresholding techniques, which can dynamically respond to changes in the system’s statistical properties, can be easily integrated into our framework with minimal modification, thereby enhancing robustness across diverse regimes. Moreover, although the method has been demonstrated on autonomous systems with fixed parameters, many real-world systems exhibit temporal variability. Extending this framework to accommodate such variations, through segmentation strategies, online learning, or adaptive models, could improve generalizability. We anticipate that convolutional neural networks (CNNs), which are capable of extracting complex temporal patterns, may still perform well under moderate variability. However, significant structural changes may require model adaptations to maintain predictive accuracy.}

\revision{Finally, while the use of machine learning for trend prediction is well established in financial time series, the application to chaotic dynamical systems presents unique challenges and opportunities. In finance, extreme value theory and models of volatility clustering have provided tools to anticipate extreme events in complex, noisy environments \cite{chatzis2018forecasting,samitas2020machine,tang2020introduction}. Drawing from this body of work, future research could explore cross-domain methods and apply them to real-time monitoring and transition detection in physical, biological, and engineered chaotic systems. As a step in this direction, we also tested our approach on the Ikeda map \cite{ott2002} and a multidimensional stochastic neuronal model \cite{boaretto2025noise}. Preliminary results showed a qualitatively similar behavior, particularly a decrease in recall as the window size and $\tau$ increase. These findings reinforce the method’s robustness across systems with distinct dynamics. It is also important to note that real-world systems often operate across different spatio-temporal scales. Accordingly, exploring multiple window sizes and spatial resolutions is essential to identify the optimal scales at which precursors and early warning signals emerge, thereby enhancing the adaptability of the proposed method to diverse application domains.}



\section*{Data availability}

The data will be made available upon request.

\section*{Acknowledgments}

E. E. N. M. would like to thanks CNPq - the Conselho Nacional de Desenvolvimento Científico e Tecnológico CNPq for the financing support for the research associated to this article; A. C. A. would like to thanks CAPES - the Coordenação de Aperfeiçoamento de Pessoal de Nível Superior, IFSP - Instituto Federal de São Paulo for the financing support. Also, A. C. A. would like to thanks the UNIFESP - postgraduate program in Computer Science at the Federal University of São Paulo;

\end{document}